\documentclass[3p,times]{elsarticle}

\usepackage{ecrc}

\volume{00}

\firstpage{1}

\journalname{Procedia Computer Science}

\runauth{}

\jid{procs}

\jnltitlelogo{Procedia Computer Science}

\usepackage{amssymb}

\usepackage[figuresright]{rotating}

\begin{document}

\begin{frontmatter}




\title{$\pi^{0}$ Azimuthal Anisotropy in Au+Au Collisions at $\sqrt{s_{NN}}=39-200$ GeV
from PHENIX: Collision Energy and Path-Length Dependence of Jet-Quenching and the Role
of Initial Geometry}


\author{Xiaoyang Gong$^{a}$ for the PHENIX Collaboration}
\address{$^{a}$Chemistry Department, Stony Brook University, Stony Brook, NY 11794, USA}

\begin{abstract}
  The azimuthal anisotropy of high $p_{T}$ particle production in heavy ion collisions is
  a sensitive probe of the jet quenching mechanism. Recent PHENIX measurements for Au+Au collisions
  at $\sqrt{s_{NN}}=200$~GeV show a $\pi^{0}$ $v_{2}$ signal that exceeds the pQCD energy loss
  calculations up to $p_{T} \sim 10$ GeV/c, challenging the traditional perturbative picture of
  the energy loss process. Here, we present an update and details of that measurement,
  as well as new high $p_T$ measurements at $\sqrt{s_{NN}}=62$ and $39$ GeV. These measurements
  not only provide an important constraint for understanding the path-length dependence of jet energy
  loss and the role of initial collision geometry, but also allow a search for the onset of jet
  quenching as $\sqrt{s_{NN}}$ is varied.
\end{abstract}




\end{frontmatter}



The discovery of the suppression of both high $p_{T}$ single particle yields \cite{PhysRevLett.88.022301}
and back-to-back jet yields in two particle correlation measurements \cite{PhysRevLett.90.082302} (``jet-quenching'')
is considered one of the most convincing piece of evidence that de-confined matter (or a quark gluon plasma [QGP]) is
created in relativistic heavy ion collisions at RHIC.
The study of jet quenching has developed into a full-fledged field with precise differential measurements
and quantitative theoretical calculations. However, our current understanding of jet quenching faces
several challenges: on the one hand, the AdS/CFT approach to describe jet-medium interactions
seems to be favored by several measurements (e.g. $R_{AA}$ for non-photonic electrons \cite{Adare:2006nq}
and high $p_{T}$ $v_{2}$ \cite{PhysRevLett.105.142301})
over the traditional pQCD based treatment; on the other hand, different models which operate
inside the pQCD framework, show large differences in their predicted quenching parameters, even
though they are embedded into the same evolving medium characterized
by 3D hydro and tuned to reproduce central $R_{AA}$ \cite{PhysRevC.79.024901}.
Precisely measured anisotropy at high $p_{T}$, can shed new insights on both challenges.

In 2007, PHENIX accumulated a large dataset of Au+Au Collisions at $\sqrt{s_{NN}}=200$GeV with several Reaction Plane (RP) detectors
which spanned a broad pseudorapidity ($\eta$) range. They included an inner ring (RXNin) ($1.0<\eta<1.5$) and an outer
ring (RXNout) ($1.5<\eta<2.8$) of a newly installed reaction plane detector (RXN), a beam beam counter (BBC) and the
muon piston calorimeter (MPC) ($3.1<\eta<3.9$). These RP detectors not only provided precise measurements of
the Event Plane (EP) but also enabled a systematic examination of non-flow effects, such as a possible jet bias.
In 2010, RHIC initiated the energy scan program and PHENIX collected several datasets
for Au+Au collisions at $\sqrt{s_{NN}}=62$ and $39$ GeV with significant statistics
with the same set of RP detectors as in 2007. The Measurements of $\pi^{0}$ $v_{2}$ from all of these datasets are discussed in the following.

The techniques used for data analysis are adopted from Ref. \cite{PhysRevLett.105.142301}. Briefly,
photons are identified and measured using the electromagnetic calorimeter (EMCal). The invariant mass
for photon pairs is sampled over events and a statistical subtraction is
implemented to remove the combinatoric background; the shape of the latter is estimated
via the widely used event-mixing technique. An integral of the subtracted invariant mass
distribution gives the $\pi^{0}$ yield. The emission angle of $\pi^{0}$ relative to the EP
$(\Delta\phi)$ is folded into $[0,\pi/2]$ (due to the symmetry in the event-averaged collision geometry;
0 for in-plane and $\pi/2$ for out-plane) and further divided into six angular bins; $\pi^{0}$ yield is
measured individually in each of the angular bins. The resulting modulation of ``$\pi^{0}$ yield vs. $\Delta\phi$''
is Fourier analyzed \cite{PhysRevC.58.1671} to obtain the $v_{2}$ term. A fit gives the so-called $v_{2}^{raw}$ and
the true corrected $v_{2}$ is the quotient $v_{2}^{raw}/\sigma_{RP}$, where $\sigma_{RP}$'s are RP resolution correction factors.
These factors characterize the dispersion of the measured EP ($\Psi$) relative to the ``true'' RP ($\Psi_{RP}$) and
are defined as $\sigma_{RP}\equiv\langle \cos2(\Psi-\Psi_{RP})\rangle$ \cite{PhysRevC.58.1671}.

%
%

\begin{figure}[htp]
  \centering
  \includegraphics[width=0.36\textwidth,height=0.36\textheight]
  {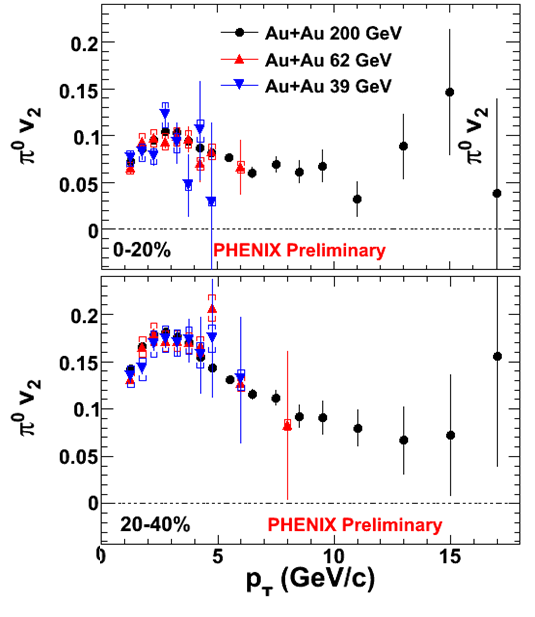}
  \includegraphics[width=0.54\textwidth,height=0.36\textheight]
  {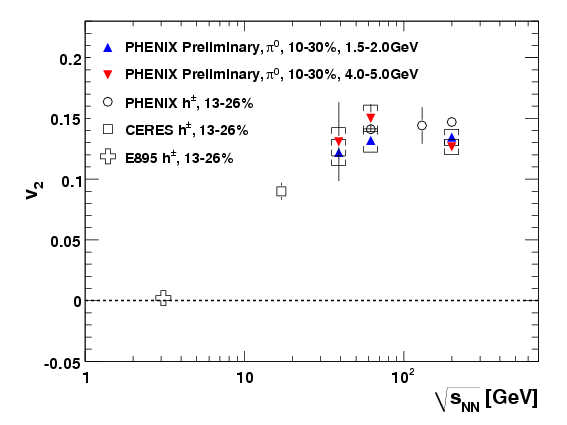}
  \caption
  {Left panel: $\pi^{0}$ $v_{2}$ vs. $p_{T}$ for 200, 62 and 39GeV for centrality $0-20\%$ and $20-40\%$.
  Right panel: a $v_{2}$ energy scan with $\pi^{0}$ and charge hadron (from \cite{PhysRevLett.94.232302}; all charge hadron
  data are of $p_{T}$ $1.5-2.0$GeV). Statistical errors are indicated by bars
  and systematic errors by square brackets.}
  \label{fig:v2}
\end{figure}

The $\pi^{0}$ $v_{2}$ results for the three collision energies are shown in the left panel of Fig-\ref{fig:v2}.
For both central ($0-20\%$) and mid-central ($20-40\%$) events the $v_{2}$ values for the respective beam energies
are quite similar. However, only the 200 GeV data has the statistical significance to probe higher $p_{T}$'s ($>6.0$GeV)
where jet fragments dominate hadron production.
Below 6 GeV $v_{2}$ is described by hydrodynamics at low $p_{T}$ ($<2.0$GeV) and recombination at intermediate $p_{T}$ ($2.0-6.0$GeV);
the good agreement of $v_{2}$'s across collision
energies indicates that $v_{2}$ is saturated in these two regions. To facilitate the investigation of $v_{2}$ saturation, the $v_{2}$ vs.
$\sqrt{s_{NN}}$ plot is presented on the right panel of Fig-\ref{fig:v2}. At low $p_{T}$, our measurements, while showing good agreement
with charge hadron results for 200 and 62GeV from \cite{PhysRevLett.94.232302}, add one more point at 39GeV. Combined with $v_{2}$
at even lower $\sqrt{s_{NN}}$ at 17 and 3GeV (measured by CERES and E895 respectively), it seems that a transition to saturation is observed.
At intermediate $p_{T}$, $v_{2}$'s are saturated down to 39GeV as well.

\begin{figure}[htp]
  \centering
  \includegraphics[width=0.6\textwidth,height=0.4\textheight]
  {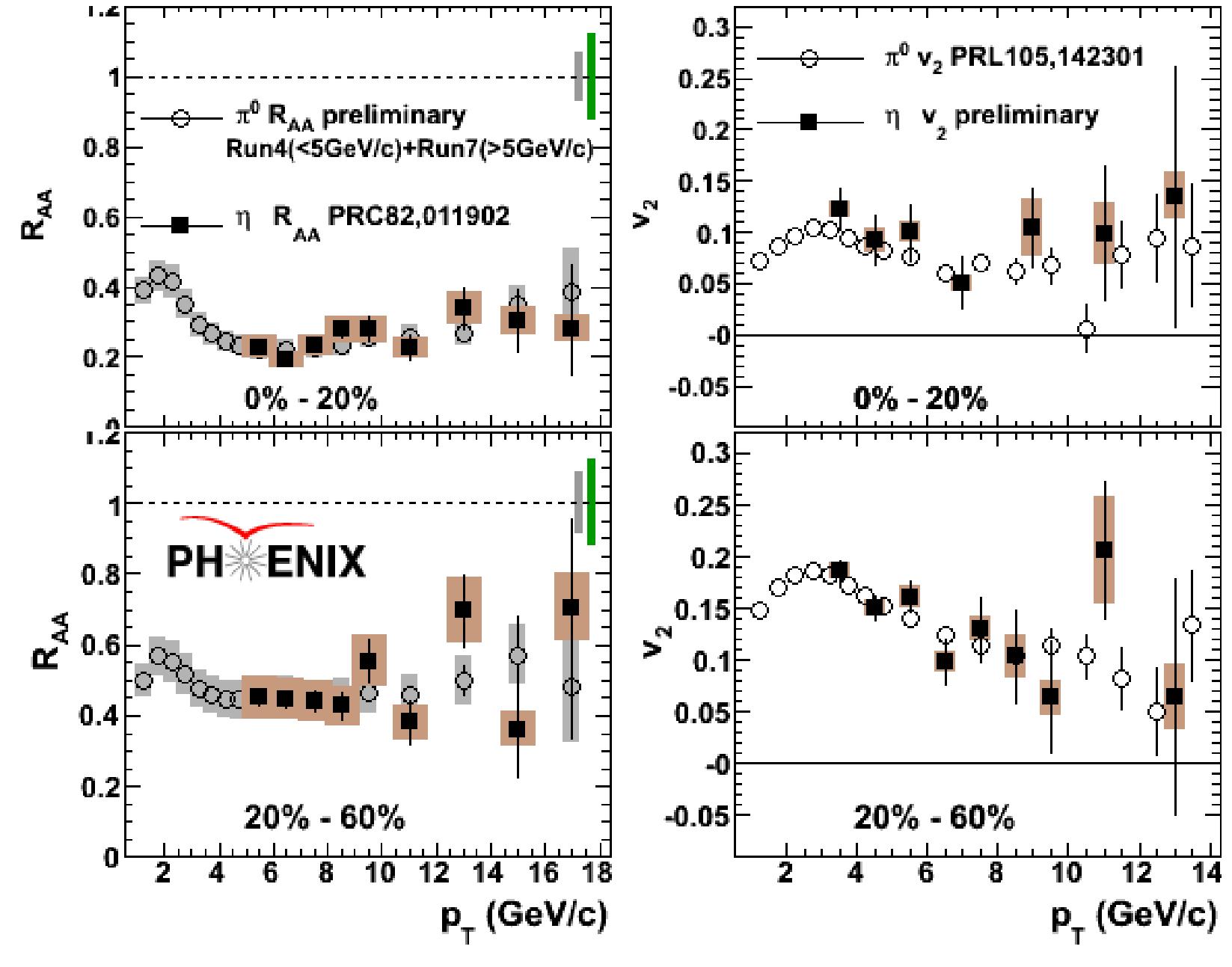}
  \caption
  {$\eta$ (black square) $R_{AA}$ (left) and $v_{2}$ (right) vs. $p_{T}$, overlaid on $\pi^{0}$ results (grep circle)
  for comparisons. Two centrality ranges are presented: $0-20\%$ (top) and $20-60\%$ (bottom).}
  \label{fig:eta}
\end{figure}

In the following we focus on the high $p_{T}$ region and discuss results from 200 GeV only. A complementary
measurement is $v_{2}$ and $R_{AA}$ for the  $\eta$ meson, which is reconstructed from two photons as well and analyzed in the
same manner as for the $\pi^{0}$. The $\eta $ has a larger mass so its yield can only be extracted for $p_{T}>4.0$GeV
in PHENIX; this is the region where jet-medium interactions play a central role. $\eta$ $v_{2}$ and $R_{AA}$ as a function
of $p_{T}$ are shown in Fig-\ref{fig:eta} for central ($0-20\%$) and mid-central ($20-60\%$) collisions;
the $\pi^{0}$ results are overlaid to aid a comparison.
The measurements show solid agreement within one standard deviation.
This observation indicates that mass and quark contents have little effects
on high $p_{T}$ $v_{2}$ and $R_{AA}$, and probably the underlying jet-quenching mechanism as well.

\begin{figure}[htp]
  \centering
  \includegraphics[width=0.64\textwidth,height=0.24\textheight]
  {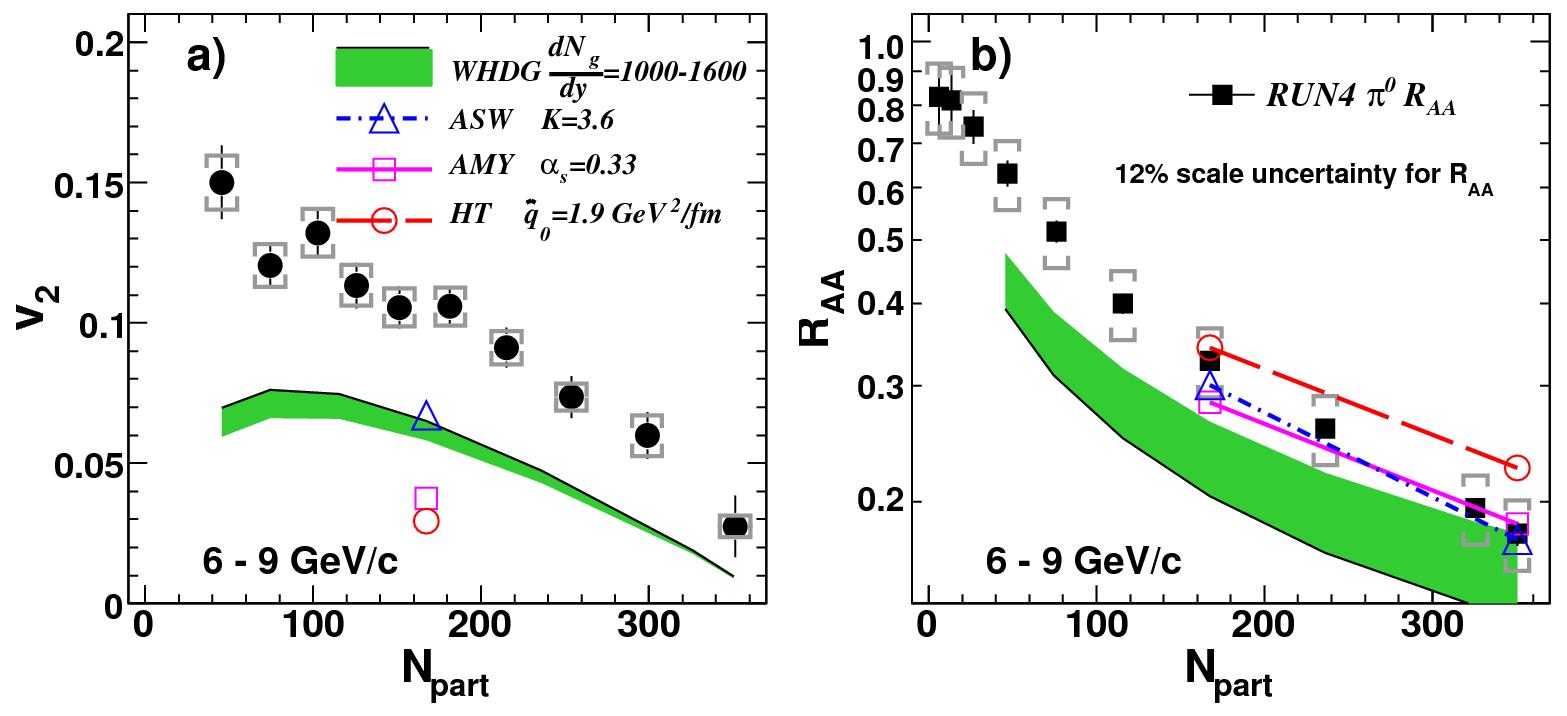}
  \caption
  {$\pi^{0}$ $v_{2}$ and $R_{AA}$ vs. $N_{part}$. Four pQCD models are plotted for comparison:
  ASW, AMY and HT from \cite{PhysRevC.79.024901}; WHDG from \cite{Wicks2007426}. The upper bound
  of $dN_{g}/dy$ in WHDG (1600) is indicated by the solid edge on the band.}
  \label{fig:v2vsNpart}
\end{figure}

Fig-\ref{fig:v2vsNpart} shows $\pi^{0}$ $v_{2}$ and $R_{AA}$ vs. $N_{part}$ (number of participants) for $6 \leq p_{T} \leq 9$ GeV/c.
The results from four popular pQCD model calculations are also plotted for comparison. Note that the parameters for these
models are tuned to match $R_{AA}$ for the most central
collisions. It is clear that while all the models successfully reproduce $R_{AA}$ for a good range of $N_{part}$
\footnote{the WHDG model undershoots the data points because it is tuned in the $p_{T}$ range $5\sim18$ GeV/c
and tends to undershoot the low $p_{T}$ end.},
they substantially under-predict $v_{2}$. Moreover, since $v_{2}$ and $R_{AA}$ are anti-correlated as indicated by
the solid edge on the WHDG band on the figure, the measurements of $v_{2}$ and $R_{AA}$ can not be simultaneously
described by these models. This discrepancy might be rooted in a combination of different effects.
In Ref. \cite{PhysRevC.82.024902} different initial collision geometries were considered; instead of the
default usage of Glauber Model geometries, the CGC type of geometry as well as its event-by-event fluctuations were
taken into account. However, the resulting enhancement of $v_{2}$ was not able to account for the whole gap.
Subsequently, a model calculation based on AdS/CFT like jet-medium interactions was tested.
The idea is that all the models above are based on pQCD;
the underestimation of $v_{2}$ may suggest that a stronger jet-medium interaction and thus, stronger path-length ($L$) dependence of jet-quenching is necessary.
The AdS/CFT type of path-length dependence is $L^{3}$, stronger than the pQCD counterpart $L^{2}$. This modification is remarkably effective;
combined with CGC geometry and event-by-event fluctuation, the reproduced $v_{2}$ matches data points well.

\begin{figure}[htp]
  \centering
  \includegraphics[width=0.80\textwidth,height=0.24\textheight]
  {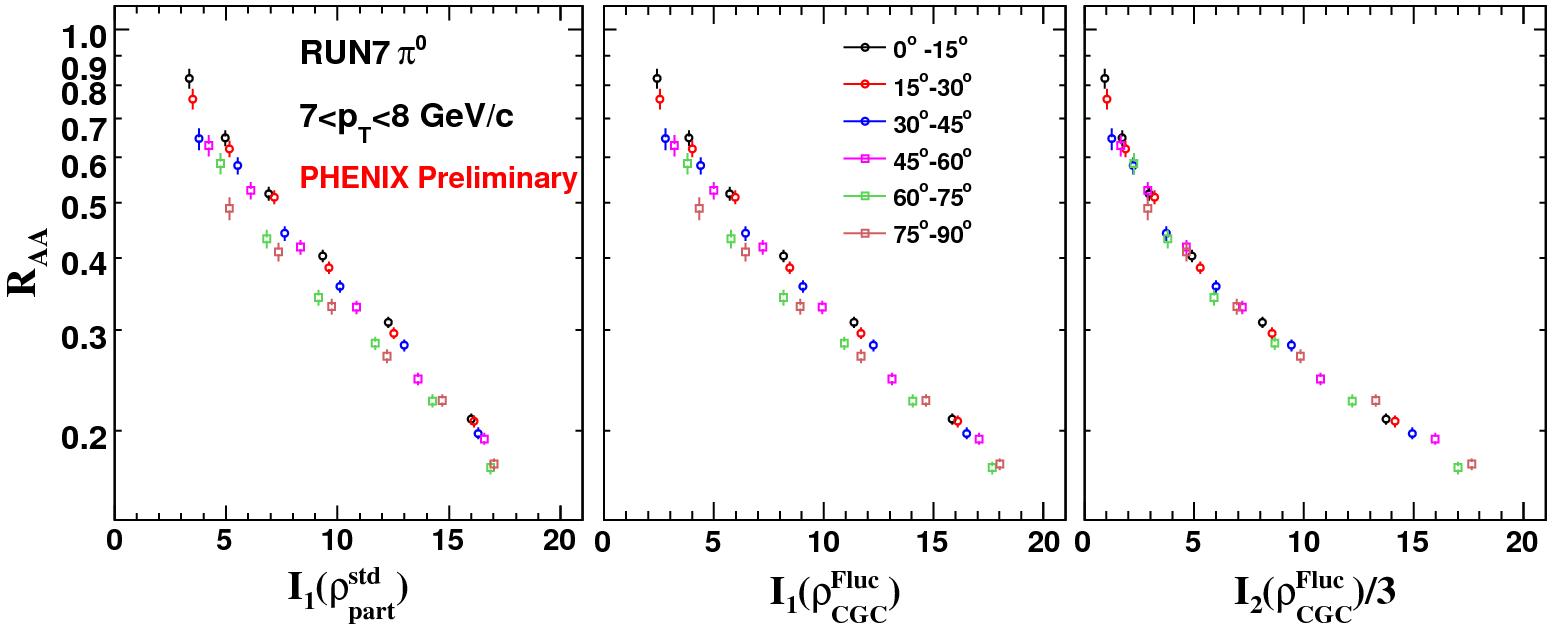}
  \caption
  {$R_{AA}(\Delta\phi,centrality)$ vs. $I_{m}$: $R_{AA}(\Delta\phi,centrality)$ is measured
  in 6 $\Delta\phi$ bins and 6 centrality ranges; $I_{m}$ is the path-length intergral and
  $m=1,2$ corresponds to pQCD and AdS/CFT like jet-medium interactions respectively.
  $\rho$ represents initial condition: $\rho^{std}_{part}$ for default Glauber geometry and
  $\rho^{Fluc}_{CGC}$ for CGC geometry with event-by-event fluctuation.}
  \label{fig:RaaScaling}
\end{figure}

To further investigate path-length dependence of jet-quenching, angular dependent $R_{AA}$ is introduced and denoted by $R_{AA}(\Delta\phi)$.
It combines the absolute suppression level $R_{AA}$ with path-length variations from in-plane to out-plane.
Fig-\ref{fig:RaaScaling} shows high $p_{T}$ ($7\sim8$ GeV/c) $\pi^{0}$ $R_{AA}(\Delta\phi)$
measured for 6 $\Delta\phi$ bins in 6 centrality bins from the most central up to $60\%$.
For each $\Delta\phi$ and centrality bin the path-length integral $I_{m}$
($m=1,2$) is calculated based on an initial geometry represented by $\rho$. On the right panel of the figure, where assumptions of
an AdS/CFT like jet-medium interaction ($I_{2}$) and CGC geometry with event-by-event fluctuation ($\rho^{Fluc}_{CGC}$) are made,
$R_{AA}$ scales remarkably with path-length integrals. This scaling suggests that
the inclusive $R_{AA}$ and $v_{2}$ for different centralities (or $N_{part}$) will be automatically and simultaneously described.

In summary, $\pi^{0}$ $v_{2}$'s for 200, 62 and 39 GeV are measured. In the low and intermediate $p_{T}$ range,
$v_{2}$ is saturated across energies.
For higher $p_T$ values we found that current pQCD models are challenged to
reproduce the $\pi^{0}$ $v_{2}$ and $R_{AA}$ simultaneously. However, an AdS/CFT-based calculation
that encodes an $L^{3}$ path-length dependence with an event-by-event fluctuating CGC geometry
is favored by the measurements.

\section*{Acknowledgments}
This work is supported by the US DOE under contract
DE-FG02-87ER40331.A008 and by the NSF under award number
PHY-1019387.





\bibliographystyle{apsrev}
\bibliography{HP2010Proceeding}







\end{document}